\documentclass[twocolumn,prl,showpacs,superscriptaddress,preprintnumbers,amsmath,amssymb]{revtex4}

\usepackage{graphicx}
\usepackage{dcolumn}
\usepackage{bm}
\usepackage{color}
\usepackage[latin1]{inputenc}
\usepackage[hang]{footmisc}
\usepackage{wasysym}
\usepackage{verbatim}

\begin{document}


\title{Experimental Evidence for Quantum Interference and Vibrationally Induced Decoherence in Single-Molecule Junctions}




\author{Stefan Ballmann}
\affiliation{Lehrstuhl f\"ur Angewandte Physik,
Friedrich-Alexander-Universit\"at Erlangen-N\"urnberg, Staudtstr.\ 7,
91058 Erlangen, Germany}
\author{Rainer H\"artle}
\affiliation{Institut f\"ur Theoretische Physik,
Friedrich-Alexander-Universit\"at Erlangen-N\"urnberg, Staudtstr.\ 7,
91058 Erlangen, Germany}
\author{Pedro B. Coto}

\altaffiliation{Instituto de Ciencia Molecular (ICMOL), Universitat de Valencia, Catedrático José Beltrán 2, 22085, E46071, Valencia, Spain}

\affiliation{Institut f\"ur Theoretische Physik,
Friedrich-Alexander-Universit\"at Erlangen-N\"urnberg, Staudtstr.\ 7,
91058 Erlangen, Germany}
\author{Marcel Mayor}
\affiliation{Institute of Nanotechnology, Karlsruhe Institute of Technology (KIT), P.O. Box 3640, 76021 Karlsruhe, Germany}
\affiliation{Department of Chemistry, University of Basel, St. Johanns-Ring 19, 4056 Basel, Switzerland}
\author{Mark Elbing}
\affiliation{Institute of Nanotechnology, Karlsruhe Institute of Technology (KIT), P.O. Box 3640, 76021 Karlsruhe, Germany}

\author{Martin R. Bryce}

\affiliation{Department of Chemistry, Durham University, Durham, DH1 3LE, United Kingdom}

\author{Michael Thoss}

\altaffiliation{Interdisziplin\"ares Zentrum f\"ur Molekulare
Materialien, Friedrich-Alexander-Universit\"at Erlangen-N\"urnberg}

\affiliation{Institut f\"ur Theoretische Physik,
Friedrich-Alexander-Universit\"at Erlangen-N\"urnberg, Staudtstr.\ 7,
91058 Erlangen, Germany}

\author{Heiko B. Weber}

\email[]{heiko.weber@physik.uni-erlangen.de}

\altaffiliation{Interdisziplin\"ares Zentrum f\"ur Molekulare
Materialien, Friedrich-Alexander-Universit\"at Erlangen-N\"urnberg}

\affiliation{Lehrstuhl f\"ur Angewandte Physik,
Friedrich-Alexander-Universit\"at Erlangen-N\"urnberg, Staudtstr.\ 7,
91058 Erlangen, Germany}

\homepage[http://www.lap.physik.uni-erlangen.de]




\date{\today}

\begin{abstract}
We analyze quantum interference and decoherence effects in single-molecule 
junctions both experimentally and theoretically by means of the mechanically 
controlled break junction technique and density-functional theory. We consider the case where interference is provided by overlapping quasi-degenerate states. Decoherence mechanisms arising 
from the electronic-vibrational coupling 
strongly affect the electrical current flowing through a single-molecule contact 
and can be controlled by temperature variation. Our findings underline the all-important relevance of vibrations for understanding 
charge transport through molecular junctions.

\end{abstract}


\maketitle

\newpage

Charge transport through single-molecule junctions \cite{cuevasscheer2010,zimbovskaya} 
is a quantum mechanical process. 
It was first treated as a purely electronic problem, combining transport concepts and 
quantum chemical calculations. Similarly, early experiments 
focused on the correlation between chemical structure and transport 
properties \cite{reichert,reichert2,elbing,tao2006}. 
Inspired by the related process of electron transfer in molecules, which is strongly dominated by 
vibration-assisted electronic phenomena, it became evident that a 
purely electronic picture is also inadequate in single-molecule transport \cite{nitzan,ho2002,galperin_2007}. 

Vibrations may appear as vibrational sidepeaks to electronic 
transitions \cite{haertle_08,haertle_09,ballmann,osorio,secker_2011}, 
and constitute a source for non-linear phenomena \cite{koch,haertle_10b} and 
bistabilities \cite{bratkovsky,galperin05}. 
In addition, vibrations may also provide a strong decoherence mechanism in molecular junctions 
where electron transport is governed by 
destructive interference effects \cite{haertle_11}. 

\begin{figure}[b!]
    \centering
        \includegraphics[width=0.72\columnwidth]{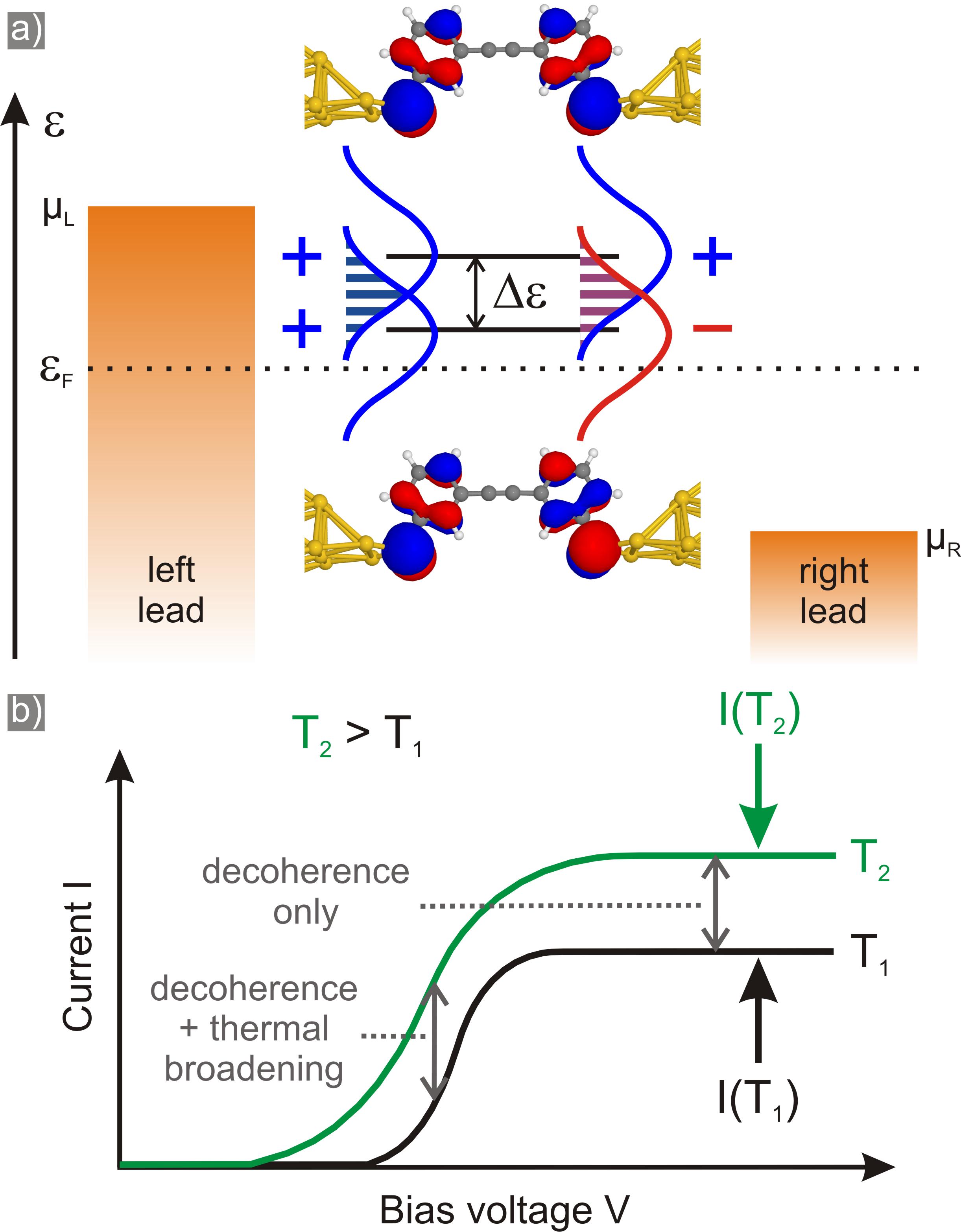}
                        \caption{(Color online) Panel (a): We consider a scenario where two molecular orbitals 
                        are close in energy ($\Delta\epsilon$) and 
                        sufficiently broadened such that they strongly overlap. We call them quasi-degenerate. In this 
                        case, interference becomes important. In particular, when two such orbitals differ in their spatial symmetry, destructive 
                        interference results in a suppression of the current flowing through the orbital pair.
                        Panel (b): In the resonant tunneling model including vibrationally induced decoherence \cite{haertle_11}, the 
                        current plateaus $I(T)$ are sensitive to vibrational excitation, which can be tuned  
                        by temperature. $I(T)$ is therefore a suitable 
                        observable to investigate vibrationally induced decoherence.
                         }
    \label{fig:depasing_T}
\end{figure}

In a purely electronic picture, destructive interference is a built-in property of many molecules commonly used in molecular 
junctions \cite{Yaliraki,Solomon,Stafford,markussen_2010,Papadopolos}. It occurs when the current is carried by 
quasi-degenerate electronic states (orbitals) which, \emph{e.g.}, differ 
in their spatial symmetry (see Fig.\ \ref{fig:depasing_T}\,a). In analogy to a double-slit experiment, these states provide different 
pathways (which are not necessarily spatially separated) for the electrons to tunnel through the molecular junction. 
Although their individual contribution to the current would be 
substantial, their phase-correct sum current can be very small (if the broadened levels overlap sufficiently, as sketched in 
Fig.\ \ref{fig:depasing_T}\,a).

Interaction with vibrations may change this result significantly \cite{haertle_11}. 
As a rule, electronic states couple very specifically to 
vibrations, in particular a different electronic-vibrational coupling of the quasi-degenerate levels is expected. Consequently, destructive 
interference between the involved pathways is quenched, which affects the electron transmission, \emph{i.e.}\ the electrical current. In more general 
terms, which-path information is provided by coupling to the vibrations.

In an experiment, vibrations are inherent to any molecular junction, in particular at room temperature. Their influence can be 
controlled by systematically varying the temperature of the junction. When starting at low temperatures $T$, the influence of vibrations 
is minimal. With increasing $T$ decoherence will continuously be tuned on \textit{via} enhanced vibrational excitation, leading to an 
effectively increased electronic-vibrational interaction.  Hence, for systems which exhibit destructive interference, the expectation is 
an increase of the electrical current with increasing $T$ (cf.\ Fig.\ \ref{fig:depasing_T}\,b).

\begin{figure}[t]
    \centering
        \includegraphics[width=0.8\columnwidth]{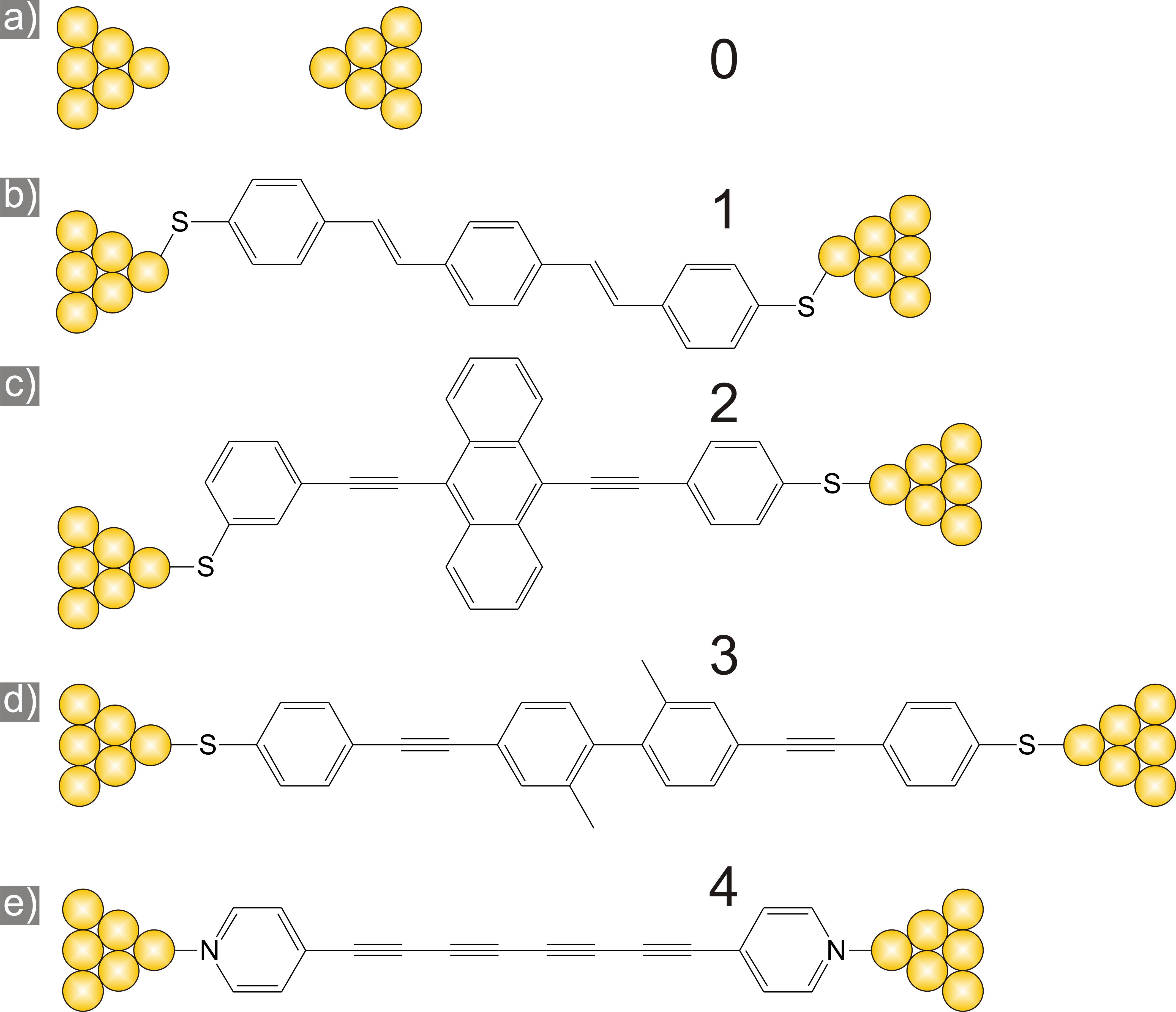}
                        \caption{(Color online) Overview of the molecular systems investigated. Panel (a): Atomically sharp 
                        gold electrodes (blind experiment). Panel (b) -- (d): Molecules \textbf{1} -- \textbf{3} with thiol anchoring groups. 
                        Panel (e): Molecule \textbf{4} with pyridine anchoring groups.}
    \label{fig:molecules}
\end{figure}

The scope of this paper is to analyze vibrationally induced decoherence in single-molecule junctions 
experimentally, accompanied by a detailed analysis of the molecular systems using density-functional theory (DFT). 
To this end, we have selected 
several molecules out of a class of rather typical molecular wires 
(conjugated, thiol-/pyridine-endcapped) \cite{danilov,elbing,wang}. The employed 
systems are depicted in Fig.\ \ref{fig:molecules}. In contrast to the molecule in the previous theoretical studies \cite{haertle_11}, which had a 
built-in L$\leftrightarrow$R symmetry, we used molecules with a reduced symmetry: in particular, 
molecule \textbf{2} has two different endgroups in meta- and para-position (cf.\ Ref.\ \cite{elbing_meta}). In molecule \textbf{3}, the conjugated system is intersected in the center by a $\sigma$ bond, and the left and right conjugated subunits are 
spatially twisted by approximately 75° with respect to each other (leading to electronic states localized either on the right or on the left end 
of the molecular rod). 
The study of systems \textbf{1}--\textbf{3} shows that L$\leftrightarrow$R symmetric molecules are only a special case of a more general 
class of systems, where quasi-degenerate states close to the Fermi energy generate vibration-sensitive interferences.
For a specific molecule, the existence of such states also depends on the molecule-lead linker group. For example, molecule \textbf{4} has the full L$\leftrightarrow$R symmetry. However, due to its pyridine anchor groups, the resulting quasi-degenerate states are located remote from the Fermi energy and, thus, do not contribute to the current. Therefore, this system provides a counter-example where it is expected 
that the considered mechanism does not affect the electrical current.

We use the mechanically controlled break junction technique at cryogenic vacuum conditions to establish single-molecule junctions 
at atomically sharp, free-standing gold tips, and record current-voltage-characteristics ($I$-$V$'s). Details of the method including 
sample processing can be found elsewhere \cite{reichert,loertscher}. 
We restricted our investigation to a temperature range of 8 to 40\,K, where the geometric structure of the junction is stable under 
temperature sweeps (see below). 
Above $40$\,K, the 
junctions exhibit 
hysteresis and/or irreversibilities of the $I$-$V$'s. 

Typical $I$-$V$ curves are sketched in Fig.\ \ref{fig:depasing_T}\,b. 
The current flowing through a molecular junction is suppressed,  
as long as the voltage $V$ is sufficiently low that the electronic levels of the junction are located  outside the energy window defined 
by the two chemical potentials $\mu_{\text{L}}$ and  $\mu_{\text{R}}$ in the left and the right 
leads, respectively ($V=(\mu_{\text{L}}-\mu_{\text{R}})/e)$. This range of bias voltages defines the 
non-resonant transport regime. If one or more electronic levels are located within the 
bias window $\left[\mu_{\text{L}},\mu_{\text{R}}\right]$, resonant transport processes occur.  

The transition is indicated by a step-like increase of the current level that remains almost 
constant for larger voltages unless another electronic state enters the bias window. This plateau has often a slightly positive slope, as quasi continously new vibrational side channels are added, which can hardly be resolved \cite{secker_2011}.  
It is on this first plateau in $I(V)$ where the analysis of the temperature dependence of the current level is carried out. It is 
particularly suited because in the resonant tunneling model (without vibrations) it is not affected by thermal 
broadening, \emph{i.e.}\ the temperature dependence of the Fermi distribution function in the leads. 

The temperature dependence of the electrical current flowing through a single-molecule junction 
has been considered before \cite{nitzan,Selzer_04,Poot_06,Choi_08,Sedghi_11}. 
However, the experimental phenomenology presented in these papers is different and 
the previous discussion focused on effects that occur at much larger temperatures ($\gtrsim150$\,K) 
and/or in the non-resonant transport regime. 

\begin{figure*}[thb]
    \centering
        \includegraphics[width=1.97\columnwidth]{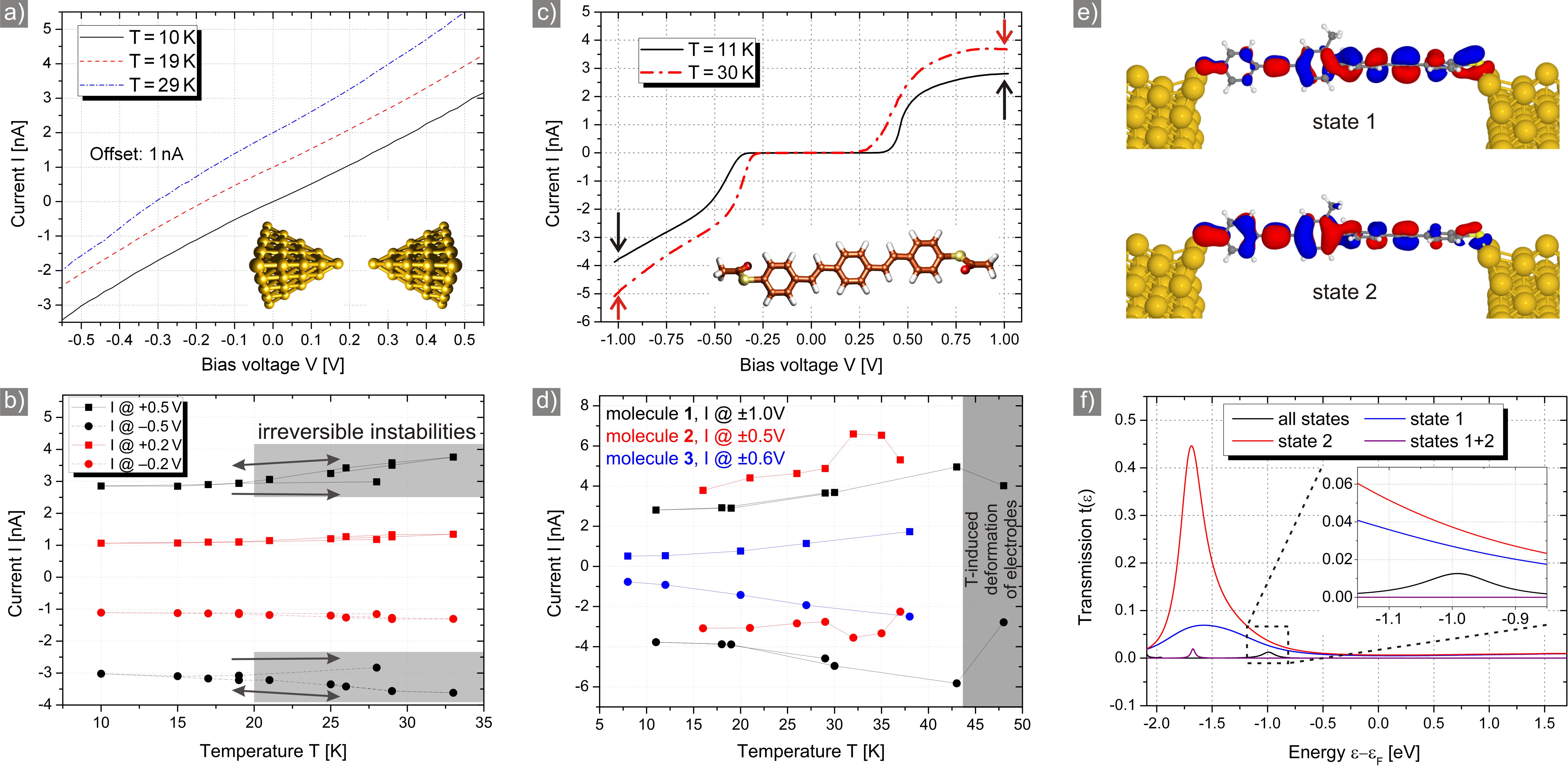}
                        \caption{(Color online)  $I$-$V$- and $I$-$T$-characteristics for gold electrodes [blind experiment, panel (a) and (b)], 
for molecule \textbf{1} [panel (c)] and for molecules \textbf{1}--\textbf{3} [panel (d)], respectively. The temperature-dependent current $I(T)$ recorded at the first plateau is the key observation indicating vibrationally induced decoherence. Panel (e): Molecular junction \textbf{3} provides 
a pair of quasi-degenerate states located next to the Fermi level. The corresponding wave functions differ in symmetry. 
Panel (f): Transmission functions for electron transport through molecular junction \textbf{3}. The individual (broadened) peaks 
associated with the quasi-degenerate states 1 and 2 at $\epsilon$\,$-$\,$\epsilon_{\text{F}}$\,$\simeq$\,$-$1.7\,eV are relatively strong, but their 
phase-correct transmission (states 1\,$+$\,2) shows only a very small peak at this energy due to destructive interference. When including all 
states this peak is shifted to $-$1.0\,eV below $\epsilon_{\text{F}}$.}
    \label{fig:au_opv3}
\end{figure*}

Before discussing the temperature dependence of single-molecule junctions, 
we present data on a blind experiment using a pair of gold electrodes which was 
treated only with solvent. The addressed question is whether the setup is sufficiently stable with respect to temperature variations. 
A temperature sweep from 10\,K to 33\,K does not show significant 
variations of the tunneling current, if the bias is low ($\lesssim $0.2\,V, see Fig.\ \ref{fig:au_opv3}\,a-b). 
At higher bias though, irreversible drifts are observed at $T$\,$>$\,22\,K. 
This small instability occurs presumably due to atomic rearrangements at high electric fields in the tip region. 
This behavior is quite common in the absence of a bridging molecule, but we 
learnt from many experiments that a bridging molecule stabilizes the junction 
and allows for stable conditions up to $\approx$\,1.5\,V at low $T$ \cite{reichert2}. We therefore conclude that in the  
temperature range considered the electrode pair provides a stable distance. 

We now investigate the temperature dependence of the current flowing through a single-molecule junction. 
As a first example, we present data obtained with molecule \textbf{1} (see Fig.\ \ref{fig:au_opv3}\,c). The 
$I$-$V$-characteristics show a typical behavior: a fairly symmetric blockade region up to $\pm$\,0.43\,V at 11\,K and 
subsequently a step-like transition to a current plateau which, however, is 
less symmetric. When measuring the very same junction at $T$\,=\,30\,K two changes are evident. 
First, the blockade region is slightly reduced to $\pm$\,0.35\,V. 
Second, the current level taken at the first plateau of the $I$-$V$-characteristics 
is significantly increased. 
This becomes evident in Fig.\ \ref{fig:au_opv3}\,d, where the current level at $\pm$\,1.0\,V 
(cf.\ arrows in Fig.\ \ref{fig:au_opv3}\,c) is plotted versus temperature. 
In contrast to the blind experiment, a clear increase of the current by about 65\% is observed, 
which remains reversible up to 
$T$\,$\approx$\,40\,K. Again, at higher temperatures, 
we observe a drift. 
This phenomenology is qualitatively reproduced for several samples.

We continue by seeking more examples for this effect, 
and investigate molecules \textbf{2} and \textbf{3}. 
For all thiol-endcapped molecular wires (\textbf{1} -- \textbf{3}), we observe qualitatively and reproducibly  
the same effect (see Fig.\ \ref{fig:au_opv3}\,d): an increase of the current with temperature that is intrinsic to the molecular junction, as inferred by the blind experiment. 
\begin{figure*}[thb]
    \centering
        \includegraphics[width=1.97\columnwidth]{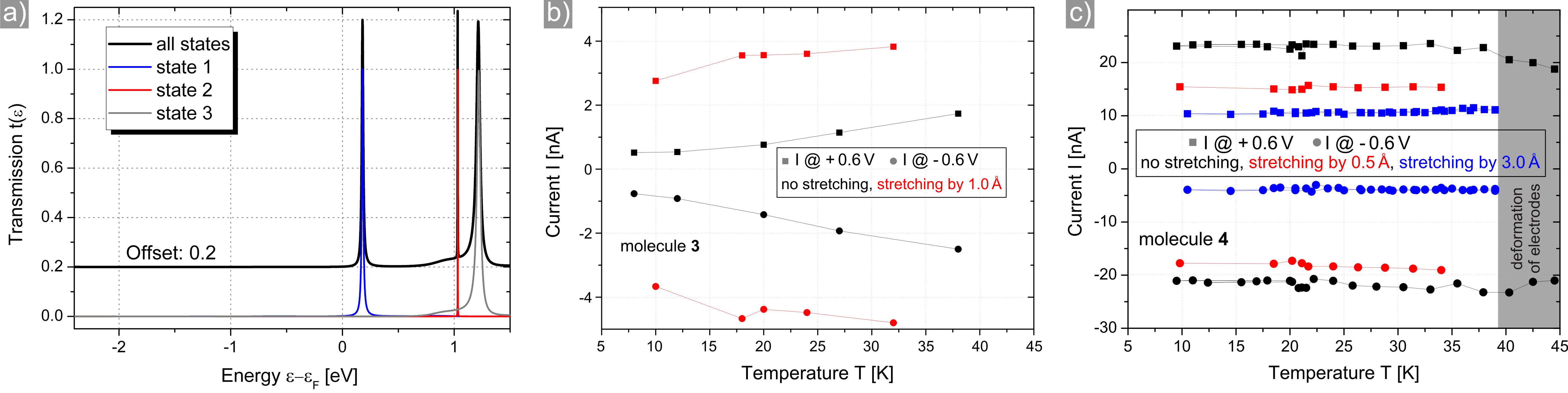}
                        \caption{(Color online) Panel (a): Transmission function for 
                        electronic transport through molecular junction \textbf{4}. The black curve is the superposition of the three 
                        individual contributions (red, blue and grey). In this molecular junction the states close to the Fermi energy are not quasi-degenerate and no sign for destructive interference occurs.
                        Panel (b) and (c): $I$-$V$- and $I$-$T$-characteristics upon stretching of molecular junctions \textbf{3} and 
                        \textbf{4}, respectively.} 
    \label{fig:symm_diode_C8_pyridin}
\end{figure*}

To rationalize the underlying mechanism, we analyze theoretically the transmission of molecular junction $\textbf{3}$,   
which shows the most pronounced temperature effect in the experiment.
To this end, we employ a model based on first-principles electronic 
structure calculations \cite{benesch_08,haertle_11} and obtain the 
respective transmission function within Landauer theory \cite{cuevasscheer2010}.  
It is depicted by the solid black line in Fig.\ \ref{fig:au_opv3}\,f. 
Peaks in this transmission function indicate that electrons can tunnel resonantly 
through the junction. Accordingly, the first plateau in the current is related to the 
peak that is closest to the Fermi level $\epsilon_{\text{F}}$. 
For our example, this is the small peak at $\epsilon$\,$-$\,$\epsilon_{\text{F}}$\,$\cong$\,$-$1\,eV 
(highlighted in the inset of this figure). 
Analysis shows that this peak results predominantly from electron transport processes through two 
electronic states. The orbitals corresponding to these two states 
are shown in Fig.\ \ref{fig:au_opv3}\,e. They represent symmetric and 
antisymmetric combinations of molecular orbitals localised in the left and 
the right part of the molecular bridge.  
If we take into account electron transport through just one of these states, we 
obtain transmission functions, which are depicted by the solid red and blue lines 
in Fig.\ \ref{fig:au_opv3}\,f. Both states 
are located relatively close in energy ($\Delta \epsilon\sim 0.02$\,eV), and are strongly broadened ($\sim0.5$\,eV). This is what we 
termed quasi-degenerate states (cf.\ Fig.\ \ref{fig:depasing_T}\,a). 
The broadening is a result of the strong coupling between the molecular bridge 
and the leads mediated by the sulfur endgroups. 
The results in Fig.\ \ref{fig:au_opv3}\,f shows that electron transport through 
these states is governed by pronounced destructive quantum interference effects \cite{haertle_11}, 
 which reduce their relatively large individual transmission function to a small peak
at $\epsilon-\epsilon_{\text{F}}\cong-1$\,eV. 
It is worth mentioning that the occurance of quasi-degenerate states is quite general 
and can be found in many molecular wires, irrespective 
of L$\leftrightarrow$R symmetry. Even the choice of electronically very different linker groups (cf.\ molecule \textbf{2}) does not suppress the effect.

As we have shown recently \cite{haertle_11}, 
electronic-vibrational coupling can strongly quench such interference effects, 
leading to substantially larger electrical currents. This quenching is enhanced 
the more strongly the vibrational degrees of freedom are excited.
A normal mode analysis of molecular junction \textbf{3} 
shows a broad distribution of vibrational modes located on the 
molecular bridge. 
Their frequencies range from $1.2$\,meV to $0.4$\,eV, including $10$ modes 
with a frequency lower than $10$\,meV (approximately $2k_{\text{B}}T_{\text{max}}$). Among these modes is the torsion (3.7\,meV) around the 
central $\sigma$ bond which is particular important for the coupling of the left and right conjugated subunit, forming states 1 and 2 in Fig.\ \ref{fig:au_opv3}\,e. Increasing the temperature in the leads 
from $8$\,K to $T_{\text{max}}$\,$=$\,40\,K leads to a substantial excitation of these modes 
and, \textit{via} the thus enhanced vibrationally induced decoherence, to a larger current level. This correlates well with the experimental 
findings (see Fig.\ \ref{fig:au_opv3}\,d). 

In order to test the correlation between a temperature dependence of the first current plateau 
and the appearance of quasi-degenerate states, we sought a counter-example. 
As the observed effect is robust even for asymmetric endgroups, we selected a completely different system: molecular wire \textbf{4}, which 
involves pyridine- instead of thiol-anchoring groups.  
The corresponding transmission function of this junction, which is  
depicted by the solid black line in Fig.\ \ref{fig:symm_diode_C8_pyridin}\,a, 
shows a number of peaks close to the Fermi level. In contrast to molecular junction \textbf{3}, each of these peaks can be associated 
with a single, non-degenerate electronic state of this junction. 
This can be inferred by comparison to the solid red, blue and gray lines 
representing the transmission probability for electron transport 
through individual states of the junction.  
This analysis shows that quantum interference effects and vibrationally induced decoherence are not 
expected to play a dominant role for electron transport through this junction. 

The experimental results confirm this prediction: the temperature dependence of the first current plateau is indeed flat for the 
pyridine-endcapped wire \textbf{4}, as displayed in Fig.\ \ref{fig:symm_diode_C8_pyridin}\,c. 
We conclude that the comparison of various molecules reveals a clear correlation: 
when quasi-degenerate states are present the first 
current plateau rises with temperature, if not, a temperature dependence is absent. 
This gives convincing evidence that decoherence due 
to coupling to vibrations plays an important role in single-molecule charge transport. 

Another independent experimental strategy to test the relevance of decoherence is to modify the coupling to the leads \textit{via} 
mechanically stretching the junction. We observed in the 
past that for thiol-endcapped molecules an \AA ngstrom-scale stretching of the molecular bridge often resulted in an increase 
of current, despite the fact that the opposite behavior is 
observed in empty tunnel junctions. We previously assumed that this modifies the transparency by altering the sulfur-gold bonding 
angle. With the given model at hand, another explanation 
occurs: when the coupling of the two quasi-degenerate levels to the leads is reduced, their mutual overlap becomes smaller and destructive 
interference effects are weakened. Experimentally this would result in (i) a larger current level  and (ii) a reduced temperature dependence.

We can test this conjecture by observing the temperature dependence upon stretching the molecular junction. 
Thereby, we assume that stretching reduces the level broadening  
due to a smaller overlap with the metal electrodes. 
This was done experimentally for molecule \textbf{3} (displaying a $T$-dependence, 
see Fig.\ \ref{fig:symm_diode_C8_pyridin}\,b) and 
\textbf{4} (displaying no $T$-dependence, see Fig.\ \ref{fig:symm_diode_C8_pyridin}\,c). 
For the thiol-endcapped molecule \textbf{3} the current 
level \textit{increases} upon stretching (in line with previous 
observations and prediction (i)) by a factor of 2-3. This can be interpreted as the reduction of destructive interference, but one may also invoke different 
explanations. For further clarification it is instructive to analyze again the temperature dependence: for molecule \textbf{3} 
the relative increase of current, $\Delta I(T)$/$I$, when changing the temperature from $8$\,K to $40$\,K is 230\%.  
This value reduces to 35\% upon stretching, in line with (ii).  
We assign this reduced 
temperature dependence to a reduction of the level broadening and a reduced overlap between the identified 
quasi-degenerate states (see Fig.\ \ref{fig:depasing_T}\,a). 

Again, the situation is different for molecule \textbf{4}, 
which has no quasi-degenerate levels close to $\epsilon_{\text{F}}$: using the 
same experimental protocol,
 the current level is slightly \textit{reduced} upon stretching, since interference plays a minor 
 role (see Fig.\ \ref{fig:symm_diode_C8_pyridin}\,c). The insensitivity to 
 temperature, however, is maintained. 
The different response upon stretching gives additional evidence to the picture that resonant electron transport through the 
thiol-endcapped molecular junctions \textbf{1}--\textbf{3} is governed by destructive interference of quasi-degenerate levels. 

To conclude, our experimental studies reveal that the electrical current flowing through a single-molecule junction 
in the resonant transport regime exhibits an intrinsic temperature dependence. By analyzing four molecules experimentally and 
theoretically, this temperature effect can be correlated with the presence of quasi-degenerate electronic states which give rise to destructive 
quantum interference. When increasing the temperature, the low-frequency modes of the junction are excited, which induces decoherence 
and thus increases the current. We expect this phenomenon to be of importance 
for a broad class of molecular wires where quasi-degenerate levels are close 
to the Fermi energy, irrespective of symmetry conditions. 
Its impact on the transport characteristics will be significantly more 
pronounced at room temperature. 
The effect underscores that vibrations play a crucial and non-trivial role 
in electronic transport through molecular junctions.


\vspace*{1em}

\begin{acknowledgments}
The work was carried out in the framework of the Cluster of Excellence \textquotedblleft Engineering of Advanced
Materials\textquotedblright\ and the SFB 953 of the DFG. Support from the EC FP7 ITN FUNMOLS is gratefully acknowledged. 
Molecule \textbf{4} was synthesized by Dr.\ Changsheng Wang (Durham University).

\end{acknowledgments}

\end{document}